# On the validity of mean-field amplitude equations for counterpropagating wavetrains

R.D. Pierce and C.E. Wayne

Dept. of Mathematics, The Pennsylvania State Univ., University Park, PA 16802

November 28, 1994

We rigorously establish the validity of the equations describing the evolution of one-dimensional long wavelength modulations of counterpropagating wavetrains for a hyperbolic model equation, namely the sine-Gordon equation. We consider both periodic amplitude functions and localized wavepackets. For the localized case, the wavetrains are completely decoupled at leading order, while in the periodic case the amplitude equations take the form of mean-field (nonlocal) Schrödinger equations rather than locally coupled partial differential equations. The origin of this weakened coupling is traced to a hidden translation symmetry in the linear problem, which is related to the existence of a characteristic frame traveling at the group velocity of each wavetrain. It is proved that solutions to the amplitude equations dominate the dynamics of the governing equations on asymptotically long time scales. While the details of the discussion are restricted to the class of model equations having a leading cubic nonlinearity, the results strongly indicate that mean-field evolution equations are generic for bimodal disturbances in dispersive systems with $O(1)$ group velocity.

# 1 Introduction

There has been considerable progress recently in demonstrating the validity of amplitude equations as approximations to the governing equations of dissipative systems [2, 4, 12]. For the most part, this work has been restricted to systems for which the initial bifurcation is either to traveling waves or steady state solutions. We concentrate on extending the results of Kirrmann, Schnieder and Mielke [4] (KSM, hereafter) to the case of counterpropagating wavetrains, of which traveling and standing waves are special cases. KSM are among the few that have considered hyperbolic systems (cf. [3] also) in addition to the more common dissipative case. We follow their derivation and prove that solutions to the amplitude equations derived below remain close to solutions of the governing equation over asymptotically long time scales. As a model we choose the governing equation to be the one-dimensional sine-Gordon (SG, hereafter) equation. However, the structure of the proof is unchanged for any equation having translation and reflection symmetries in both space and time as well as a leading cubic nonlinearity. The translation symmetries imply that the imposition of periodic boundary conditions results in a system which is O(2)-symmetric. When this property is combined with the fact that the group velocity is of $O(1)$, asymptotic scaling arguments indicate that the correct modulation equations are of the mean-field type [5, 6, 7, 9, 10]. However, in related systems other arguments have lead to the conclusion that the coupling should be local [8]. In an effort to resolve this question, we first derive the mean-field nonlinear Schrödinger equations (MFNLS, hereafter) as the asymptotically exact amplitude equations for the SG equation. We then rigorously prove that the approximate solution represented by the amplitudes remains asymptotically close to a full solution to the governing equation on time scales of $O(\epsilon^{-2})$. Here $\epsilon$ is a small parameter measuring the size of the disturbance. In addition, we show that similar estimates hold both for solutions in $H^1(R)$, and also in spaces of periodic functions. We conclude with a discussion of how our results are relevant to studies of wave propagation in physical systems.

## 2 The Amplitude Equations

We look for solutions $u(x,t) : R \times [0,\infty) \to R$ to the SG equation,

$$\partial_t^2 u = \partial_x^2 u - \sin u \ . \tag{1}$$

If we restrict attention to solutions of the *linearized* problem with amplitude of $O(\epsilon)$ and periodic in $x$ with period $\frac{2\pi}{k}$, we have:

$$u_L = \epsilon A^+ e^{i(kx-\omega t)} + \epsilon A^- e^{i(kx+\omega t)} + cc \ , \tag{2}$$

where the $A^{\pm}$ are $O(1)$ constants and $cc$ denotes the conjugate of all preceding terms. The frequency $\omega$ is related to the wavenumber $k$ by the dispersion relation

$$\omega^2 = k^2 + 1 \ . \tag{3}$$

$u_L$ is a valid solution to (1) in the limit that $\epsilon \to 0$. Note that (1) has a translational symmetry in both $x$ and $t$. Each wavetrain in $u_L$ breaks these symmetries, but each preserves a spatio-temporal symmetry in which translations in space are equivalent to translations in time. Together they break the remaining symmetry because the spatial translations are in opposite directions for a given temporal displacement, but the wavetrains are decoupled in the linear problem and so each retains its symmetry properties independent of the other.

We continue to examine the linear case, but we now relax the periodic boundary conditions, allowing for a continuous wavenumber spectrum. The spatio-temporal translation symmetry of each wavetrain is broken by modulation of the envelopes in space and time. In this case, it appears that all of the symmetries of the problem are now broken. However, modulations of the wave envelope with infinitesimal wavenumber propagate at the group velocity, so that each envelope is time independent in the characteristic frame moving at its group velocity, $c_g = \frac{k}{\omega}$ for the SG equation. Consequently, there is a hidden spatio-temporal translation symmetry associated with each wave envelope, for which translations in space correspond to translations in time. It is important to note that this hidden symmetry is a property of the *linear* problem, but it is a symmetry of the amplitudes and does not correspond to a symmetry of the full modulated linear solution. The net result of this argument is to establish that if we wish to describe modulated linear waves, then the amplitudes should be functions of the characteristic variables $A^\pm = A^\pm(x \mp c_g t)$.

For finite but small $\epsilon$, the nonlinear terms introduce perturbations to the linear problem. This causes the amplitudes to vary in time and space, and it also causes the wavetrains to interact. However, since the perturbed solution must converge to the linear solution in the limit that $\epsilon \to 0$, the amplitudes may vary only on scales which are asymptotically long in the limit, i.e., $A^\pm = A^\pm(\epsilon t, \epsilon x, \epsilon^2 t, \cdots)$, and interactions between the wavetrains must take place at higher order in $\epsilon$. Due to the hidden translation symmetry, the amplitudes cannot have arbitrary dependence on the slow variables $\epsilon x$ and $\epsilon t$ but must instead depend only on the slow characteristic variables $\chi^\pm \equiv \epsilon(x \mp c_g t)$ at leading order. Moreover, terms which break this hidden symmetry can only enter at higher order in $\epsilon$. Note that we can only make this argument because the hidden symmetry is a property of the *linear* problem. The approximate solution which is expected to be valid for some finite range of $\epsilon$ should then have the form:

$$u_A = \epsilon(A^+(\chi^+, \tau) + \epsilon B^+(\chi^+, \chi^-, \tau))e^{i(kx-\omega t)} + \epsilon(A^-(\chi^-, \tau) + \epsilon B^-(\chi^+, \chi^-, \tau))e^{i(kx+\omega t)}$$
$$+ cc , \qquad (4)$$

where $\tau \equiv \epsilon^2 t$ is a super-slow time scale. The $B^\pm$ are the leading order terms which break the hidden symmetry, and, like the $A^\pm$, they represent long wavelength modulations of the linear solution. As noted above, they must enter the expansion at $O(\epsilon^2)$ at the earliest because only nonlinear effects can break the hidden symmetry. There are no second harmonics because the nonlinearity enters at $O(\epsilon^3)$. We have not included a super-slow spatial scale, $\xi \equiv \epsilon^2 x$, because our results prove that it is not necessary in order to capture the leading order behavior for time scales of $O(\epsilon^{-2})$.

Following KSM, we introduce an "improved" approximation, from which we will derive the amplitude equations:

$$v_A = \epsilon(A^+ + \epsilon B^+)e^{i(kx-\omega t)} + \epsilon(A^- + \epsilon B^-)e^{i(kx+\omega t)}$$

$$+\epsilon^3 \Big\{ \frac{1}{6(9k^2 - 9\omega^2 + 1)} (A^{+3} e^{3i(kx-\omega t)} + A^{-3} e^{3i(kx+\omega t)})$$

$$+ \frac{1}{2(9k^2 - \omega^2 + 1)} (A^{+2} A^{-} e^{i(3kx-\omega t)} + A^{+} A^{-2} e^{i(3kx+\omega t)})$$

$$+ \frac{1}{2(k^2 - 9\omega^2 + 1)} (A^{+2} \overline{A^{-}} e^{i(kx-3\omega t)} + \overline{A^{+}} A^{-2} e^{i(kx+3\omega t)}) \Big\} + cc \,, \qquad (5)$$

where the overbar denotes conjugation. The cubic terms are designed to cancel the nonresonant terms resulting from the expansion of the nonlinearity in the governing equation. For the current problem, these higher harmonics are nonresonant for all $k \neq 0$, but for other problems this is not always true and the resonant harmonics must explicitly be introduced to the expansion at lower order. We next compute the amount by which $v_A$ fails to be a solution by substituting it into the governing equation and calculating the "residuum":

$$\begin{aligned}
\rho(\epsilon, x, t) &\equiv \partial_t^2 v_A - \partial_x^2 v_A + \sin v_A \\
&= \epsilon^3 \Big[ -2i\omega \partial_\tau A^{+} - 4ik \partial_{\chi^-} B^{+} + (c_g^2 - 1) \partial_{\chi^+}^2 A^{+} - \frac{1}{2} |A^{+}|^2 A^{+} - |A^{-}|^2 A^{+} \Big] e^{i(kx-\omega t)} \\
&\quad + \epsilon^3 \Big[ 2i\omega \partial_\tau A^{-} - 4ik \partial_{\chi^+} B^{-} + (c_g^2 - 1) \partial_{\chi^-}^2 A^{-} - \frac{1}{2} |A^{-}|^2 A^{-} - |A^{+}|^2 A^{-} \Big] e^{i(kx-\omega t)} \\
&\quad + cc + O(\epsilon^4) \,. \qquad (6)
\end{aligned}$$

The quantities in the square brackets are essentially the amplitude equations, but the evolution of both the first and second order amplitudes cannot be resolved from the two equations. The correct way to proceed is to separate the terms which depend on the counterpropagating characteristic variables into separate equations. This is equivalent to interpreting the amplitude equations as inhomogeneous equations for the $B^{\pm}$ and applying the consequent solvability condition (cf. Section 4 for further discussion).

$$2i\omega \partial_\tau A^{+} = (c_g^2 - 1) \partial_{\chi^+}^2 A^{+} - (\frac{1}{2} |A^{+}|^2 + \sigma^{-}) A^{+} \qquad (7)$$

$$-2i\omega \partial_\tau A^{-} = (c_g^2 - 1) \partial_{\chi^-}^2 A^{-} - (\frac{1}{2} |A^{-}|^2 + \sigma^{+}) A^{-} \qquad (8)$$

$$4ik \partial_{\chi^-} B^{+} = -(|A^{-}|^2 - \sigma^{-}) A^{+} \qquad (9)$$

$$4ik \partial_{\chi^+} B^{-} = -(|A^{+}|^2 - \sigma^{+}) A^{-} \qquad (10)$$

The $\sigma^{\pm}$ are functions of the super-slow time $\tau$ only, and their definitions depend on the function space in which we seek solutions. If solutions are sought in the form of localized wave packets, then

$$\sigma^{\pm}(\tau) \equiv 0 \ . \tag{11}$$

If the amplitudes are forced to be periodic of period $P_1$, the $\sigma^{\pm}$ are defined as

$$\sigma^{\pm}(\tau) \equiv \frac{1}{P_1} \int_0^{P_1} |A^{\pm}(s,\tau)|^2 \, ds \ . \tag{12}$$

The terms $B^{\pm}$ have now decoupled from the amplitudes, which must evolve according to the MFNLS equations, (7) and (8). The solutions of (9) and (10) for $B^{\pm}$ may be written explicitly:

$$B^{\pm}(\chi^+, \chi^-, \tau) = \frac{i}{4k} A^{\pm}(\chi^{\pm}, \tau) \int_\ell^{\chi^{\mp}} (|A^{\mp}(s,\tau)|^2 - \sigma^{\mp}(\tau)) \, ds \ , \tag{13}$$

where $\ell \equiv -\infty$ in the whole-line case and $\ell \equiv 0$ in the periodic case. Note that the $B^{\pm}$ remain of $O(1)$ regardless of whether the $A^{\pm}$ are localized or periodic. Note also that the $B^{\pm}$ are determined by (9) and (10) only up to the addition of an arbitrary function $\gamma^{\pm}(\chi^{\pm}, \tau)$. In writing (13), we have set $\gamma^{\pm} \equiv 0$. A different choice of functions may be physically relevant in some cases, but this presents no additional complications provided that the $\gamma^{\pm}(\cdot, \tau)$ are bounded in the appropriate norm.

## 3  Estimates on the Validity of the Amplitude Equations

While the preceeding arguments are convincing, we now show that they can, in fact, be made mathematically rigorous. We begin by discussing the case of localized wavetrains, since the discussion in this case parallels that of KSM. Let $u(x,t)$ be the solution of the SG equation. As KSM point out, the natural space in which to consider this problem is $(u, \partial_t u) \in Y = H^1(R) \times L^2(R)$. The global well-posedness of SG in this space is a standard result. Similarly, by the results of [1] (section 6), one has global well-posedness of (8) and (9) (with $\sigma^{\pm} = 0$) in $H^1(R)$. Finally, consider $B^{\pm}$ which solve (9) and (10) (and for which we have an explicit representation in (13)). From (13) one has immediately that

$$|B^{\pm}(\chi^+, \chi^-, \tau)| \leq \frac{1}{4|k|} |A^{\pm}(\chi^{\pm}, \tau)| \int_{-\infty}^{\chi^{\mp}} |A^{\mp}(\xi, \tau)|^2 \, d\xi \leq \frac{1}{4|k|} |A^{\pm}(\chi^{\pm}, \tau)| \|A^{\mp}\|_{L^2}^2 \ . \tag{14}$$

As an immediate consequence of this estimate we have:

**Lemma 1**

$$\int_{-\infty}^{\infty} |B^{\pm}(\epsilon x + \epsilon c_g t, \epsilon x - \epsilon c_g t, \epsilon^2 t)|^2 \, dx \leq \frac{1}{16k^2\epsilon} \|A^{\pm}(\cdot, \epsilon^2 t)\|_{L^2}^2 \|A^{\mp}(\cdot, \epsilon^2 t)\|_{L^2}^4 \ , \tag{15}$$

*while*

$$\|B^{\pm}(\cdot, \cdot, \epsilon^2 t)\|_{L^{\infty}} \leq \frac{1}{4|k|} \|A^{\pm}(\cdot, \epsilon^2 t)\|_{L^{\infty}} \|A^{\mp}\|_{L^2}^2 \ . \tag{16}$$

**Remark 1** *By the Sobolev inequality,* $\|A^{\pm}\|_{L^{\infty}}^2 \leq 2\|A^{\pm}\|_{L^2}\|A^{\pm}\|_{H^1}$.

Note that these estimates allow us to bound the $L^2$ and $L^{\infty}$ norms of our approximate solution $v_A$.

**Corollary 1** *Let* $\|A\|_{L^2} = \max(\|A^+\|_{L^2}, \|A^-\|_{L^2})$ *and similarly for* $\|A\|_{H^1}$. *Then for $\epsilon$ sufficiently small, there exists a constant $c(k,\omega)$ such that* $\|v_A\|_{H^1} \leq c(k,\omega)\epsilon^{1/2}\|A\|_{H^1}$.

**Proof:** The estimates of the lemma allow us to bound

$$\left| \int A^{\pm}(\epsilon x \pm c_g t) B^{\pm}(\epsilon x + c_g t, \epsilon x - c_g t, \epsilon^2 t) \, dx \right| \leq \frac{1}{4|k|\epsilon} \|A^{\pm}\|_{L^2}^2 \|A^{\mp}\|_{L^2}^2 \ . \tag{17}$$

Thus, taking into account the scaling of the spatial variable, we find

$$\|v_A(\cdot, \tau)\|_{L^2}^2 \leq c(k,\omega) \left( \epsilon \|A\|_{L^2}^2 + \epsilon^2 \|A\|_{L^2}^4 + \epsilon^3 \|A\|_{L^2}^3 \|A\|_{H^1} + \epsilon^4 \|A\|_{H^1}^6 \right) \ . \tag{18}$$

For $\epsilon$ sufficiently small, this is bounded by $c(k,\omega)\epsilon\|A\|_{H^1}^2$. The estimate on $\|v_A\|_{H^1}$ is similar and we don't repeat it. □

Note further that as a corollary of these estimates on $A^{\pm}$ and $B^{\pm}$ and the fact that they satisfy (7) - (10), we obtain the following estimates of the residuum (6):

**Corollary 2** *Suppose that $A^{\pm}$ are solutions of (7) and (8), for which $\partial_\tau^\alpha \partial_{\chi_\pm}^\beta A^{\pm} \in C([0,T_0], L^2(R))$ for $\alpha + \beta \leq 2$. Then there exists a constant $c > 0$, depending on $k$, $\omega$, and the norm of $A^{\pm}$ and its derivatives of order 2 or less, such that for all $0 \leq t \leq T_0\epsilon^{-2}$,*

$$\|\rho(\epsilon, t, \cdot)\|_{L^{\infty}} \leq c\epsilon^4 \tag{19}$$

$$\|\rho(\epsilon, t, \cdot)\|_{L^2} \leq c\epsilon^{7/2} \tag{20}$$

$$\|\rho(\epsilon, t, \cdot)\|_{H^1} \leq c\epsilon^{7/2} \ . \tag{21}$$

**Remark 2** *If one assumes that the initial data $A^{\pm}(x,0)$ of (7) and (8) is sufficiently smooth, it is straightforward to show that $\partial_\tau^\alpha \partial_{\chi_\pm}^\beta A^{\pm} \in C([0,T_0], L^2(R))$ using ideas like those used by Cazenave ([1], section 5) to prove well-posedness in $H^2$.*

We now estimate the error $R(x,t) \equiv \epsilon^{-3/2}(u(x,t) - v_A(x,t))$, where $u(x,t)$ is a solution of the SG equation. Given the estimates on $v_A$, we can follow KSM almost word for word through these estimates. Note first that

$$\partial_t^2 R = \partial_x^2 R - R + a(\epsilon,t) R + \epsilon^{-3/2} N(\epsilon,t,R) - \epsilon^{-3/2} \rho(\epsilon,t) \,, \tag{22}$$

where $a(\epsilon,t) = (1 - \cos(v_A))$, and $N(\epsilon,t,R) = \epsilon^3 \frac{1}{2} \sin(v_A + \theta \epsilon^{3/2} R) R^2$, for some $|\theta| \leq 1$.

**Remark 3** *If we had chosen to consider a more general nonlinear equation than the SG equation, say $\partial_t^2 u = \partial_x^2 u - u + g(u)$, the only change in this expression would be that $a(\epsilon,t) = g'(v_A)$, and $N(\epsilon,t,R) = \epsilon^3 \frac{1}{2} g''(v_A + \theta \epsilon^{3/2} R) R^2$.*

As KSM note, if we rewrite (22) as a first order system in terms of the variables $(R,S) = (R, \partial_t R)$, then the linear evolution

$$\partial_t \begin{pmatrix} R \\ S \end{pmatrix} = \begin{pmatrix} S \\ \partial_x^2 R - R \end{pmatrix} \tag{23}$$

preserves the norm $\|(R,S)\|_Y^2 = \int_{-\infty}^{\infty} (\partial_x R)^2 + R^2 + S^2\, dx$. Since $\|v_A\|_{L^\infty} \leq C\epsilon$, $|a(\epsilon,t)| \leq C\epsilon^2$ for all $t$, while $N(\epsilon,t,R)$ is a smooth function from $Y$ to $L^2$ with $\|N\|_{L^2} \leq \epsilon^4 \|R\|_{H^1}^2$. Thus, we can rewrite (22) as the integral equation:

$$\begin{pmatrix} R(t) \\ S(t) \end{pmatrix} = G(t) \begin{pmatrix} R(0) \\ S(0) \end{pmatrix} + \int_0^t G(s-t) \begin{pmatrix} 0 \\ a(\epsilon,t) R(s) + \epsilon^{-3/2} N(\epsilon,t,R) - \epsilon^{-3/2} \rho(\epsilon,s) \end{pmatrix} ds \,, \tag{24}$$

where $G(t)$ is the semigroup associated with the linear equation (23). Using the fact that $G$ preserves the $Y$ norm, standard estimates imply that

$$\begin{aligned}
\|(R(t),S(t))\|_Y &\leq \|(R(0),S(0))\|_Y + \epsilon^{-3/2} \sup_{0 \leq s \leq t} \|\rho(\epsilon,s)\|_{H^1} t \\
&\quad + \int_0^t \{C_1 \epsilon^2 \|(R(s),S(s))\|_Y + C_2 \epsilon^{5/2} \|(R(s),S(s))\|_Y^2\}\, ds \,,
\end{aligned} \tag{25}$$

where we have used the fact that if $\|R(s)\|_{H^1} \leq \tilde{C}$, then $\|\sin(v_A + \theta\epsilon^{3/2}R)\|_{L^\infty} \leq C_2\epsilon$, for $\epsilon$ sufficiently small. Next presume that we can choose our initial amplitudes $A^\pm(\chi^\pm, 0)$ so that $\|u(x,0) - v_A(x,0)\|_{H^1} \leq C\epsilon^{3/2}$, and $\|\partial_t u(x,0) - \partial_t v_A(x,0)\|_{L^2} \leq C\epsilon^{3/2}$. Thus, we can assume that $\|(R(0), S(0))\|_Y \leq C_1$. Similarly, using our estimate on $\|\rho(\epsilon, s)\|_{H^1}$, from Corollary 2, we see that $\epsilon^{-3/2} \sup_{0 \leq s \leq t} \|\rho(\epsilon, s)\|_{H^1} t \leq CT_0$. Thus, we can apply Gronwall's inequality as in KSM to conclude:

**Theorem 1** *Suppose that $A^\pm$ are solutions of (7) and (8), for which $\partial_\tau^\alpha \partial_{\chi^\pm}^\beta A^\pm \in C([0, T_0], L^2(R))$ for $\alpha + \beta \leq 2$. Then there exists $\epsilon_0$ and $C_0$ greater than zero, such that if $0 \leq \epsilon < \epsilon_0$ the solution of (24) satisfies $\|(R(t), S(t))\|_Y \leq C_0$, for all $0 \leq t \leq T_0 \epsilon^{-2}$.*

From this one immediately concludes:

**Corollary 3** *Under the hypotheses of the previous theorem, there exist positive constants $\epsilon_0$ and $C_0$ such that if $\epsilon < \epsilon_0$, then*

$$\|u(\cdot, t) - v_A(\cdot, t)\|_{H^1} \leq C_0 \epsilon^{3/2} , \tag{26}$$

*for all $0 \leq t \leq T_0 \epsilon^{-2}$. By Sobolev's inequality, one immediately has*

$$\|u(\cdot, t) - v_A(\cdot, t)\|_{L^\infty} \leq C_0 \epsilon^{3/2} , \tag{27}$$

*for the same time interval.*

We now turn to a consideration of the problem with periodic boundary conditions. We consider solutions $u(x, t)$ which are periodic with period $P_\epsilon$, where $P_\epsilon$ and $\frac{2\pi}{k}$ are commensurate. This implies (because of the scaling of the spatial variables) that the amplitude functions $A^\pm$ and $B^\pm$ are periodic with period $\epsilon P_\epsilon$, and hence we choose $P_\epsilon = \epsilon^{-1} P_1$ and $P_1$ is an $O(1)$ constant. Define a norm

$$\|u(\cdot, t)\|_{Y_{per}(\epsilon)}^2 = P_\epsilon^{-1} \int_0^{P_\epsilon} \{u^2 + (\partial_x u)^2 + (\partial_t u)^2\} \, dx . \tag{28}$$

We also define the norms $\|u(\cdot, t)\|_{L^2_{per}(\epsilon)} = P_\epsilon^{-1} \int_0^{P_\epsilon} u^2 \, dx$, and by analogy $H^1_{per}(\epsilon)$.

In like fashion we will define the $H^1_{per}$ norm for $A^\pm$ by

$$\|A^\pm(\cdot,\tau)\|^2_{H^1_{per}} = P_1^{-1} \int_0^{P_1} \{|A^\pm(\chi^\pm,\tau)|^2 + |\partial_{\chi^\pm} A^\pm(\chi^\pm,\tau)|^2\} d\chi^\pm \ , \tag{29}$$

and analogously for $L^2_{per}$.

**Remark 4** *The well-posedness of (7) and (8) follow immediately if we note that $\sigma^\pm(\tau) \equiv \|A^\pm(\cdot,\tau)\|^2_{L^2_{per}} = \|A^\pm(\cdot,0)\|^2_{L^2_{per}}$ are independent of $\tau$ and define $A^\pm = \tilde{A}^\pm \exp(\frac{\pm i\sigma^\mp \tau}{2\omega})$. The $\tilde{A}^\pm$ then satisfy uncoupled NLS equations.*

Now consider $B^\pm$ defined by

$$B^\pm(\chi^+,\chi^-,\tau) = \frac{i}{4k} A^\pm(\chi^\pm,\tau) \int_0^{\chi^\mp} (|A^\mp(\xi,\tau)|^2 - \sigma^\mp(\tau)) d\xi \ . \tag{30}$$

Note that the periodicity of $A^\pm$ and the presence of the counterterm $\sigma^\mp$ ensure that $B^\pm$ is periodic in both variables as required. Similarly, one can repeat the estimates leading up to Lemma 1 and one finds:

**Lemma 2**

$$P_\epsilon^{-1} \int_0^{P_\epsilon} |B^\pm(\epsilon x + \epsilon c_g t, \epsilon x - \epsilon c_g t, \epsilon^2 t)|^2 \, dx \ \leq \ \frac{1}{16k^2} \|A^\pm(\cdot,\epsilon^2 t)\|^2_{L^2_{per(\epsilon)}} \|A^\pm(\cdot,\epsilon^2 t)\|^4_{L^2_{per(\epsilon)}} \tag{31}$$

$$\|B^\pm(\cdot,\cdot,\epsilon^2 t)\|_{L^\infty} \ \leq \ \frac{1}{4|k|} \|A^\pm(\cdot,\epsilon^2 t)\|_{L^\infty} \|A^\mp(\cdot,\epsilon^2 t)\|^2_{L^2_{per(\epsilon)}} \ . \tag{32}$$

One can now proceed to estimate our approximate solution $v_A$ exactly as we did in the whole-line case. We find

**Lemma 3** *Let $\|A\|_{H^1_{per}} = \max(\|A^+\|_{H^1_{per}}, \|A^-\|_{H^1_{per}})$, and similarly for $\|A\|_{L^\infty}$. Then*

$$P_\epsilon^{-1} \int_0^{P_\epsilon} (|v_A(x,t)|^2 + |\partial_x v_A(x,t)|^2) \, dx \ \leq \ C(k,\omega)\{(\epsilon \|A\|_{H^1_{per}} + \epsilon^2 \|A\|_{H^1_{per}})^2$$
$$+\epsilon^4 \|A\|^2_{L^\infty} \|A\|^2_{H^1_{per}} (1 + \|A\|^2_{H^1_{per}} + \|A\|^2_{L^\infty})\} \ . \tag{33}$$

**Remark 5** *There is a slight difference here between the periodic and whole-line problems which concerns the estimate of $\|A\|_{L^\infty}$. In order to estimate this, choose $x_0^\pm$ to be points such that $|A^\pm(x_0^\pm)|^2 = \sigma^\pm$. Then $|A^\pm(x)|^2 - \sigma^\pm = 2\int_{x_0^\pm}^x A^\pm(\xi)\partial_{\chi^\pm}A^\pm(\xi)\,d\xi$, so we conclude that $\|A^\pm\|_{L^\infty}^2 \leq \sigma^\pm + 2P_1 \|A\|_{H_{per}^1}^2$.*

The important point is that although somewhat more complicated than than the corresponding estimate in the whole-line situation, the right hand side of this inequality is still bounded, independent of $\epsilon$ for all $0 \leq t \leq T_0$.

In like fashion we can derive estimates on the residuum (6). We find:

**Corollary 4** *Suppose that $A^\pm$ are solutions of (7) and (8), for which $\partial_\tau^\alpha \partial_{\chi^\pm}^\beta A^\pm \in C([0, T_0], L_{per}^2)$ for $\alpha + \beta \leq 2$. Then there exists a constant $c > 0$, depending on $k$, $\omega$, and the norm of $A^\pm$ and its derivatives of order 2 or less, such that for all $0 \leq t \leq T_0 \epsilon^{-2}$,*

$$\|\rho(\epsilon, t, \cdot)\|_{L^\infty} \leq c\epsilon^4 \tag{34}$$

$$\|\rho(\epsilon, t, \cdot)\|_{H_{per}^1} \leq c\epsilon^4. \tag{35}$$

Once again, we emphasize that these estimates can be chosen independent of time, for $0 \leq t \leq T_0 \epsilon^{-2}$.

We complete our analysis of the periodic case by estimating the error $R_{per}(x, t) = \epsilon^{-2}(u(x,t) - v_A(x,t))$. (Note the change in scaling with respect to the whole-line case. This results from the change in the way the norms scale with $\epsilon$ in the periodic case.)

$$\partial_t^2 R_{per} = \partial_x^2 R_{per} - R_{per} + a(\epsilon, t)R_{per} + \epsilon^{-2}N_{per}(\epsilon, t, R_{per}) - \epsilon^{-2}\rho, \tag{36}$$

where as before $a(\epsilon, t) = (1 - \cos(v_A))$ and $N_{per}(\epsilon, t, R_{per}) = \epsilon^4 \frac{1}{2}\sin(v_A + \epsilon^2\theta R_{per})R_{per}^2$. As before we note that the semi-group $G_{per}$ associated with the linearized equations (36) (subject to periodic boundary conditions) preserves the norm $\|(R_{per}, S_{per})\|_{Y_{per}}^2$. Hence, as before we can estimate (36) by rewriting it as an integral equation and we find:

$$\|(R_{per}(t), S_{per}(t))\|_{Y_{per}} \leq \|(R_{per}(0), S_{per}(0))\|_{Y_{per}} + \epsilon^{-2}\sup_{0 \leq s \leq t}\|\rho(\epsilon, s, \cdot)\|_{H_{per}^1(\epsilon)}t$$

$$+ \int_0^t \{C\epsilon^2 \|(R_{per}(s), S_{per}(s))\|_{Y_{per}} + C\epsilon^3 \|(R_{per}(s), S_{per}(s))\|^2_{Y_{per}}\} ds \ . \tag{37}$$

Just as in the whole-line case, we choose our initial amplitudes so that $\|(R_{per}(0), S_{per}(0))\|_{Y_{per}} \leq \tilde{C}$. Thus by another application of Gronwall's lemma, we conclude:

**Theorem 2** *Suppose that $A^\pm$ are solutions of (7) and (8), for which $\partial_\tau^\alpha \partial_{\chi^\pm}^\beta A^\pm \in C([0, T_0], L^2_{per})$ for $\alpha + \beta \leq 2$. Then there exists $\epsilon_0$ and $C_0$ greater than zero, such that if $0 \leq \epsilon < \epsilon_0$ the solution of (37) satisfies $\|(R(t), S(t))\|_{Y_{per}} \leq C_0$, for all $0 \leq t \leq T_0 \epsilon^{-2}$.*

As before, we can immediately conclude from this that $\|u(\cdot, t) - v_A(\cdot, t)\|_{H^1_{per}(\epsilon)} \leq C_0 \epsilon^2$, for all $0 \leq t \leq T_0 \epsilon^{-2}$. But then if we choose $x_0$ so that $(u(x_0, t) - v_A(x_0, t))^2 = P_\epsilon^{-1} \int_0^{P_\epsilon} (u(x, t) - v_A(x, t))^2 \, dx \leq C_0 \epsilon^4$, we have

$$\begin{aligned}
(u(x, t) - v_A(x, t))^2 &= (u(x_0, t) - v_A(x_0, t))^2 \\
&\quad + 2 \int_{x_0}^x (\partial_x u(\xi, t) - \partial_x v_A(\xi, t))(u(\xi, t) - v_A(\xi, t)) \, d\xi \\
&\leq C_0 \epsilon^4 + \int_{x_0}^x \{(\partial_x u(\xi, t) - \partial_x v_A(\xi, t))^2 + (u(\xi, t) - v_A(\xi, t))^2\} \, d\xi \\
&\leq C_0 \epsilon^4 + 2 P_\epsilon \|u - v_A\|^2_{H^1_{per}(\epsilon)} \leq C_0 \epsilon^4 + 2 P_1 \epsilon^3 \ .
\end{aligned} \tag{38}$$

Thus, we have proven the following corollary:

**Corollary 5** *Under the hypotheses of the preceeding theorem, there exist positive constants $\epsilon_0$, $C_1$, such that if $0 \leq \epsilon < \epsilon_0$ then the solution of (36) satisfies $\|u(\cdot, t) - v_A(\cdot, t)\|_{L^\infty} \leq C_1 \epsilon^{3/2}$ for all $0 \leq t \leq T_0 \epsilon^{-2}$.*

## 4 Discussion

We begin by considering alternative approaches to eliminate the $O(\epsilon^3)$ terms from the residuum (6), and describe why we feel our method is the most appropriate. The quantities in the square brackets are recognizable as NLS equations with a cubic cross term and, effectively, a slow time derivative of the symmetry breaking $B^\pm$ terms. These derivatives may well be regarded as the "problem" terms. It is tempting to argue that they might be neglected because they are higher order terms, either by breaking the hidden symmetry at leading order or simply setting the $B^\pm$ to zero. In the former case, one might try to define $A^\pm \equiv A^\pm(\epsilon t, \epsilon x, \cdots)$, but the hidden symmetry is restored at $O(\epsilon^2)$ via the evolution equations

$$\partial_{\chi^\mp} A^\pm = 0 \ . \tag{39}$$

The right-hand side of these equations are identically zero due to the absence of a quadratic nonlinearity in the governing equation. If the alternative approach is taken and the $B^\pm$ are set to zero, the amplitude equations take the form of locally coupled NLS equations (i.e., (7) and (8) with $\sigma^\pm \equiv |A^\pm|^2$). Considering the evolution equation for $A^+$, it is independent of the characteristic variable $\chi^-$ except for a parametric variation of the term $|A^-|^2 A^+$, and similarly for the $A^-$ equation. Grouping terms depending on $\chi^+$ and $\chi^-$ on opposite sides of the equations, it is apparent that the quantities $|A^\pm|^2$ depend only on $\tau$ and not on the characteristic variables $\chi^\pm$, at least as long as the amplitudes do not have zeroes. Consequently, the amplitudes take the form $A^\pm \equiv a^\pm(\tau) e^{i\phi^\pm(\chi^\pm, \tau)}$, and the locally coupled NLS equations are found to exclude the possibility of amplitude modulation. While one must entertain the possibility that this is the physically relevant description, there are two compelling arguments against it. Mathematically, the process of setting the $B^\pm$ to zero contains the *unjustified assumption* that the terms which break the hidden symmetry (and destroy the existence of the characteristic frames) appear at $O(\epsilon^3)$ and not at the first opportunity as one generically expects. Physically, the exclusion of amplitude modulation is contrary to the existence of the Benjamin-Feir instability, which is a well-documented feature of wavetrains in physical systems.

The results presented above generalize those of KSM in several ways. Most importantly, we have addressed the case of counterpropagating wavetrains and the case of periodic boundary conditions. The details of the proof restrict our results to governing equations having a leading cubic nonlinearity, thereby excluding most systems of hydrodynamical interest. It is likely that these results will hold for systems having quadratic nonlinearities, provided that the possibility of resonant triads is excluded, but it should be noted that it is *not* a trivial exercise to prove this rigorously [11]. Moreover, we have shown that the mean-field terms in the amplitude equations originate with a hidden symmetry of the *linear* problem. Consequently, it is unlikely that the structure of the nonlinearity will have a substantial effect on the nature of the mean-field coupling of the wavetrains.

It has also been demonstrated recently that the asymptotically exact equations governing small-amplitude broad-band disturbances (the so-called AEZ equations) are similar to the well-known Zakharov equations [13], but the AEZ equations differ in that they contain a time average [9]. At the same time it was shown that the averaging is responsible for generating the mean-field terms in, for instance, the MFNLS equations derived here; the original Zakharov equations without the average yield locally coupled NLS equations. Consequently, we interpret the present results as strong, though not conclusive, evidence for the generic nature of mean-field coupling for both broad-band and nearly monochromatic small-amplitude disturbances.

## Acknowledgements

We would like to acknowledge valuable discussions with Edgar Knobloch. One of us (RDP) would also like to thank Andrew Bernoff, Diane Henderson and Bernard Matkowsky for helpful remarks. While this work was being completed, one of us (CEW) was a visitor at the Mittag-Leffler Institute whose hospitality is gratefully acknowledged. This work was supported by NSF grants DMS 9257456 and DMS 9203359.